\documentclass{article}

\usepackage{graphicx} 
\usepackage{authblk}
\usepackage{caption} 
\usepackage{float}
\usepackage[version=3]{mhchem}
\RequirePackage{amsmath,amsfonts,amssymb}
\RequirePackage{mathptmx} 

\usepackage[scaled]{helvet}

\usepackage{xcolor}

\newcommand{\make}[1]{\textbf{\textcolor{red}{#1}}}
\newcommand{\loss}[1]{\textbf{\textcolor{blue}{#1}}}

\RequirePackage{authblk}
\RequirePackage[left=2cm,%
    right=2cm,%
    top=2.25cm,%
    bottom=2.25cm,%
    headheight=12pt,%
    letterpaper]{geometry}%
    
\RequirePackage[labelfont={bf,sf},%
    labelsep=period,%
    justification=raggedright]{caption}

\RequirePackage[colorlinks=true, allcolors=blue]{hyperref}

\usepackage[resetlabels,labeled]{multibib}

\RequirePackage[superscript,biblabel,nomove]{cite}

\usepackage{geometry}
\newgeometry{vmargin={15mm}, hmargin={26mm,26mm}} 

\title{A dry Venusian interior constrained by atmospheric chemistry}
\date{November 2024}

\author[1]{Tereza Constantinou}
\author[1,2]{Oliver Shorttle}
\author[3]{Paul B. Rimmer}

\affil[1]{Institute of Astronomy, University of Cambridge, Madingley Road, Cambridge, CB3 0HA, UK}
\affil[2]{Department of Earth Sciences, University of Cambridge, Downing Street, Cambridge CB2 3EQ, UK}
\affil[3]{Cavendish Astrophysics, University of Cambridge, JJ Thomson Avenue, Cambridge CB3 0HE, UK}

\begin{document}

\maketitle

\begin{abstract}

Venus's climatic history provides powerful constraint on the location of the inner-edge of the liquid-water habitable zone. However, two very different histories of water on Venus have been proposed: one where Venus had a temperate climate for billions of years, with surface liquid water, and the other where a hot early Venus was never able to condense surface liquid water. Here we offer a novel constraint on Venus's climate history by inferring the water content of its interior.  By calculating the present rate of atmospheric destruction of \ce{H2O}, \ce{CO2} and \ce{OCS}, which must be restored by volcanism to maintain atmospheric stability, we show Venus's interior is dry. Venusian volcanic gases have at most a 6\% water mole fraction, substantially drier than terrestrial magmas degassed at similar conditions. The dry interior is consistent with Venus ending its magma ocean epoch desiccated and thereafter having had a long-lived dry surface. Volcanic resupply to Venus's atmosphere therefore indicates that the planet has never been `liquid-water' habitable.

\end{abstract}

\section*{Introduction}\label{sec:intro}

Venus's geodynamic and climatic history is uncertain, particularly whether the surface ever hosted habitable conditions.  At its heart, this is a question of whether liquid water oceans have ever existed on Venus. Unlike for Mars, where spectacular mapping of the surface reveals its sculpting by water\cite{carr1987water}, Venus's global resurfacing at-least $\sim$0.3\,billion years ago has obscured much of its ancient geological history\cite{Strom1994}.  One observation in support of water, in some form, having existed on Venus is its large deuterium/hydrogen (D/H) ratio of $150\,\pm30$ times that of Earth\cite{donhue}.  This observation has been used to infer that Venus once had a surficial water reservoir of similar mass to Earth's oceans \cite{raymond2017}: equivalent to up to 500\,m of global water depth on Venus\cite{donahue1997venus, Warren}. Alternatively, the observed D/H ratio could be a result of steady state supply: in this scenario water is continuously delivered through cometary material or volcanic outgassing, balancing loss through H escape \cite{grinspoon1988cometary}. Critically, the atmospheric D/H ratio alone does not constrain the existence of liquid water on Venus, regardless of what it implies about the quantity or source of water. Hence, the past surface and climatic conditions of Venus are still unknown.

\begin{figure}[htbp]
	\includegraphics[width=\columnwidth]{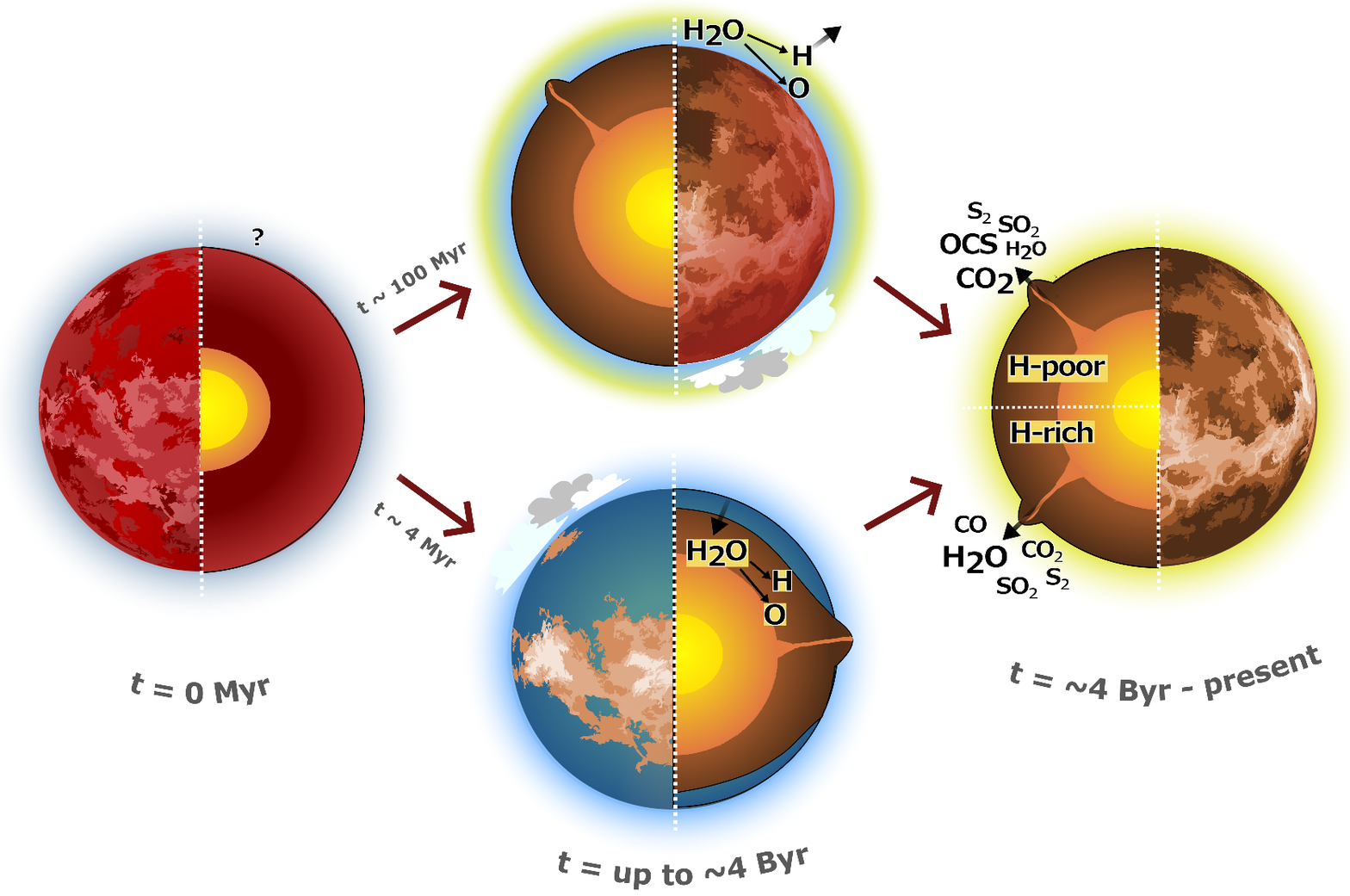}
 \caption{ \textbf{The dichotomous climate pathways proposed for Venus.}  Venus emerges from its formation in a magma ocean phase ($\text{t}=0\,\text{Myr}$).  From this point Venus may follow one of two possible pathways, each ultimately converging on the planet's present climate state, but leaving a mantle of different composition: an H-poor interior results from the ``dry Venus'' (upper) path, and an H-rich interior from the ``temperate and wet Venus'' (lower) path. In the H-rich interior scenario, volcanic gases are enriched in \ce{H2O}, while in the H-poor case, degassing of \ce{H2O} is minimal, with a greater release of S and C species. See Introduction for a detailed description of climatic paths, and Section \nameref{sec:interior} for an explanation of how \ce{H2O} oceans are imprinted in the planetary interior.}
 \label{fig:venus_drawing}
\end{figure}

In the absence of definitive observations, climate modelling has been crucial in informing our picture of early Venus, from which two contrasting paradigms emerge: ``temperate and wet Venus'' and ``dry Venus'' (Fig. \ref{fig:venus_drawing}). In the temperate scenario, Venus had an Earth-like past: a lasting period of temperate climate followed the planet's formation, with shallow oceans (e.g., refs. \citenum{1988Kasting, GrinspoonBullock2007}). General Circulation Models (GCMs) demonstrate such a state can be maintained if Venus is initialised cool, with liquid water at its surface, by water clouds that form on its day-side \cite{Way2016,Way2020}. A cloudy dayside and clear night-side reduce radiation absorption and enhance efficient re-radiation to space, keeping surface temperatures cool enough for water condensation. From this habitable state, Venus transitioned to its present \ce{CO2} greenhouse during the ``great climate transition'', likely having occurred by massive outgassing of \ce{CO2} and \ce{SO2} during large igneous province volcanism \cite{BULLOCK200119,Way2020}.

In contrast to this is the ``dry Venus'' scenario.  Here, Venus became desiccated early in its evolution due to its slow magma ocean solidification over $\sim100\,$Myr\cite{hamano2013emergence}. The long-lived steam atmosphere would have enabled dissociation and loss of water through intense hydrodynamic escape of H\cite{GILLMANN2009503}, which would have extended throughout the whole planetary interior\cite{hamano2013emergence}.  With subsequent volcanic outgassing, Venus would have evolved towards its current \ce{CO2}-\ce{N2}-dominated atmosphere. This scenario is also supported by a GCM, where an initially hot Venus without surface liquid water leads to night-side water cloud formation\cite{Turbet2021}.  These night-side clouds have a strong net warming effect (blocking the radiation of day-side heat to space), and prevent surface water condensation even at moderate insolations \cite{Turbet2021}. Such a scenario predicts that an ocean-free early Venusian surface will remain ocean-free thereafter. Any water preserved in the hot atmosphere after magma ocean solidification may then have been slowly lost over the planet's history\cite{selsis2023cool}.

These dichotomous predictions of Venus's evolution are illustrated in Fig. \ref{fig:venus_drawing}.  The essence of the different outcome of GCMs is whether the models are initialised hot (in which case the planet cools inefficiently\cite{Turbet2021}), or cold with an ocean (in which case it can remain temperate for geological timescales\cite{Way2016}).  A key question then is whether Venus could cool sufficiently following the high-energy conditions of planetary assembly (Fig. \ref{fig:venus_drawing}, left) to allow liquid water to form on its surface (the lower path of Fig. \ref{fig:venus_drawing}), or whether a hot-start forces early water loss (the upper path of Fig. \ref{fig:venus_drawing}) and a permanently dry Venusian surface\cite{Gillmann2022}. 

To distinguish between the dichotomous climate scenarios demonstrated in Fig.\ref{fig:venus_drawing}, we present a novel approach informed by observations of Venus's current atmospheric chemistry.   We calculate what the net composition of volcanic resupply to Venus's atmosphere must be in order to maintain atmospheric steady state at the present-day.  This, in turn, constrains how wet the planet's interior is, which serves as the source of volcanism. Therefore we can constrain the planet's internal water inventory, which is distinct between the two dichotomous climate histories. 

\section*{Signs of climate in the Interior}\label{sec:interior}

To determine if water once existed on Venus's surface in the form of oceans --- indicating traditional liquid-water habitability --- we investigate whether Venus has a water-rich interior at the present-day. We suggest that a habitable past will be associated with Venus's present interior being water-rich, and a dry past with Venus’s present interior being dry. 

The majority of Venus’s water inventory was likely delivered during the main accretion phase, and not during late accretion\cite{gillmann2020dry}. This makes Venus's magma ocean stage key in setting the initial water inventory and distribution on the planet. Slow magma ocean solidification significantly depletes a planet's water inventory: depletion can occur down to less than a tenth of Earth's ocean mass even when starting with a water budget ten times larger\cite{hamano2013emergence}. Following magma ocean crystallisation, a significant proportion of the water \emph{left} partitions into, and \emph{remains} within, the planet's mantle\cite{dorn2021hidden, bower2022retention, Miyazaki, luo2024majority}.

The above discussion indicates that if Venus's water inventory had been depleted during the magma ocean stage, then its surface would always have been dry thereafter. In this scenario, volcanism would have provided a one-way path for water out of the planet's interior. Water in the atmosphere would then dissociate into H and O, and the H would hydrodynamically escape the planetary system\cite{Turbet2021}. 

If Venus was once habitable (i.e., had surface oceans), however, we would expect a water-rich system, necessitating rapid magma ocean solidification, minimising H loss to space following water condensation\cite{hamano2013emergence}. The ``dry Venus'' scenario is therefore linked to an H-poor interior, whereas the ``temperate and wet Venus'' scenario is linked to an H-rich interior, drawing a direct connection between the current interior water reservoir, and the past behaviour of the water in the atmosphere (Fig. \ref{fig:venus_drawing}).

If Venus retained a significant water reservoir following its magma ocean phase, then much of that water should still be present in its interior. The maximum time window between modern-day Venus and a past Venusian climate that could have supported a surface oceans is 3\,Gyr\cite{Way2020}. Stagnant lid outgassing rates are likely\cite{armann2012simulating} inefficient, and suggest a maximum of 0.4 Earth-oceans degassed over this time span\cite{GUIMOND2021106788}; less than half the likely water inventory of a temperate Venus\cite{Way2016}. A temperate Venus would also have diminished water degassing due to high-pressure eruption under surface oceans\cite{KrissansenTotton_2021}, further increasing the likelihood of a wet past being reflected by a wet mantle. Noble gas tracers of volcanic degassing rates point directly to the planet having had time-integrated inefficient degassing compared to Earth \cite{kaula1999constraints}: perhaps only 24\% of the radiogenic $^{40}$Ar in Venus' mantle has been released\cite{o2015thermal}. Limited Ar loss points to a similarly inefficient loss of water, preserving evidence of a wet past in its interior.

Information on Venus's present interior is afforded through its volcanism. Volcanism on rocky planets is overwhelmingly supplied by partial melting of their hot convecting mantles.  Elements such as H are typically incompatible during these melting processes\cite{Aubaud2004,HIRSCHMANN201238}, and are therefore efficiently transported towards the lower pressure surface where they degas into the atmosphere. The majority of the mantle water inventory is in the upper mantle\cite{guimond2023mantle}, where it can depress the melting temperature of mantle rocks\cite{Katz}, increasing the likelihood volcanism will sample wet mantle domains. Thus the composition of the present day volcanic gases on Venus, i.e., the proportion of water in the volcanic gas, irrespective of the total rate or volume of volcanism, links back to the water content of the interior, and critically, to the planet's climatic past.

\section*{Sources and sinks of the Venusian atmosphere}\label{sec:atmo}

\subsection*{Photochemical losses}\label{sec:atmo_losses}
\begin{figure*}[htbp]\centering
	\includegraphics[scale=0.6]{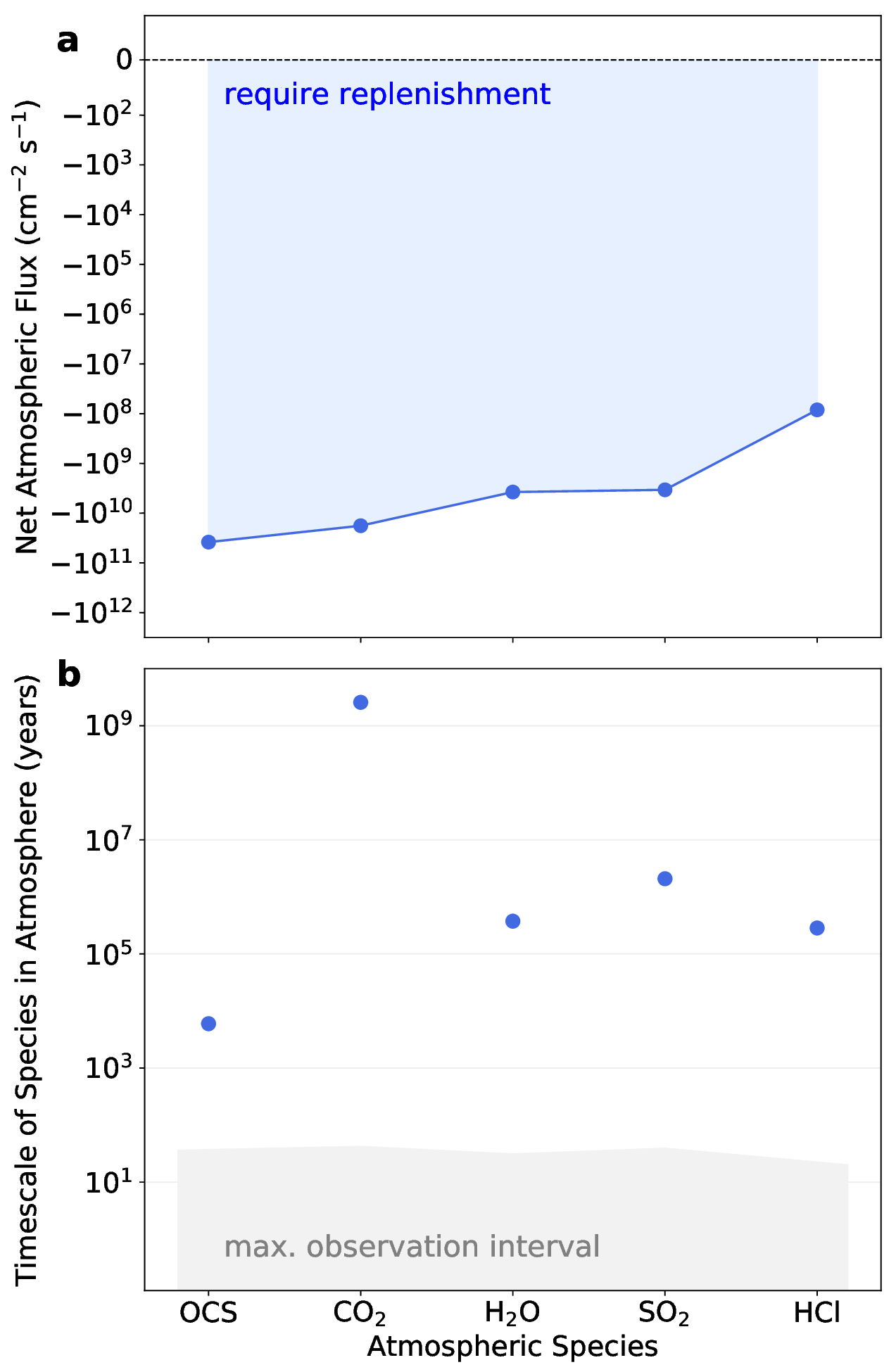}
 \caption{ \textbf{Dominant destructive fluxes of species in Venus's atmosphere. a}, Calculated net atmospheric flux for the 5 species with the highest rates of destruction (blue line). Species undergoing atmospheric destruction necessitate an equivalent flux for replenishment to maintain atmospheric steady-state (blue shaded region). \textbf{b,} Timescales for species in the atmosphere, determined by the duration it would take for the abundance of the species undergoing destruction to decrease by a factor of $e$ (blue circles). The baseline of lower limits of stability of species in the atmosphere, determined by the largest time interval across observational measurement (grey shaded area). Data sources\protect\cite{ANDERSON1968, OyamaPIONEER, KRASNOPOLSKY2008377, 2008Marcq, Greaves2021}. }
 \label{fig:flux_time}
\end{figure*}

We calculate atmospheric production and destruction rates of gases using a chemical-kinetic network that reproduces the observed modern atmosphere of Venus \cite{rimmer2021hydroxide}. Figure \ref{fig:flux_time} shows the species with the highest rates of destruction in the atmosphere, and their effective chemical timescales. For the atmosphere to be in steady state, the total flux of a species $X$ in the atmosphere, $\Delta\Phi_X$ (cm$\sf{^{-2}}$ s$\sf{^{-1}}$), should be zero (\nameref{sec:methods}). The gaseous species \ce{OCS}, \ce{CO2}, and \ce{H2O} exhibit the highest rates of net destruction in the atmosphere, accompanied by correspondingly high production rates of other C-O-H-S species, as dictated by the conservation of elemental abundance (see Supplementary Text 1 and Supplementary Fig. 1).

The primary reaction pathways dominating over the atmospheric column are (key species are highlighted: \loss{blue} is loss, \make{red} is production):
\begin{align}
\loss{\ce{OCS}} + hv &\rightarrow \ce{CO} + \ce{S}, \label{eqn:chem_3}\\
\loss{\ce{CO2}} + hv &\rightarrow \ce{CO} + \ce{O}. \label{eqn:chem_4}
\end{align}
These photochemical dissociation reactions dominate at the top of the atmosphere ($> 90 \, {\rm km}$ from the surface).

Within the clouds, the formation of sulfuric acid from \ce{SO3} and \ce{H2O}, and its degradation, provide the ``seeds'' for cloud formation. However, these reactions are in near-perfect balance, and therefore contribute negligible net production or destruction of either \ce{H2O} or \ce{SO3}.

Closer to the surface ($< 45 \, {\rm km}$), the following reactions dominate:
\begin{align}
\loss{\ce{CO2}} + \ce{HN} &\rightarrow \ce{CO} + \ce{HNO},\label{eqn:chem_5} \\
\ce{CO} + \ce{HNO} &\rightarrow \make{\ce{CO2}} + \ce{HN}, \label{eqn:chem_6} \\
\loss{\ce{OCS}} + \ce{S} &\rightarrow \ce{CO} + \ce{S2}, \label{eqn:chem_7}\\
\ce{CO} + \ce{S} &\rightarrow \make{\ce{OCS}}. \label{eqn:chem_8}
\end{align}
And water chemistry is set by:
\begin{align}
\loss{\ce{H2O}} + \ce{S2O} &\rightarrow \ce{H2S} + \ce{SO2}, \\
\ce{H2SO3} + \ce{H2O} &\rightarrow 2\make{\ce{H2O}} + \ce{SO2}.\label{eqn:chem_10}
\end{align}
The net effect of the above reactions results in the set of dominant atmospheric destruction fluxes shown in Figure \ref{fig:flux_time}. Supplementary Text 2 and Supplementary Figs. 2--4 expand further on reaction rates.

Measurements of the atmospheric abundance of species over the twentieth and twenty-first centuries place lower limits on the timescales for the evolution of the planet's atmosphere (grey region Fig. \ref{fig:flux_time}).  While observations of Venus's atmospheric chemical abundances have been made at different altitudes and lack consistent methodology, they align with our self-consistent atmospheric calculations for Venus\cite{rimmer2021hydroxide}, and are ultimately consistent (within error) with Venus having a broadly constant atmospheric composition over this period\cite{2023NatCoHelbert}. 

The key abundance ratios of species being atmospherically destroyed are OCS/\ce{H2O} and \ce{CO2}/\ce{H2O}. These ratios are both $>1$, which if directly resupplied in these abundances would indicate a very dry source of gas to the atmosphere; hence, a very dry magma if volcanism provided the restorative flux.  

However, before making the connection to volcanism and the interior of Venus, it is important to note that balancing photochemical destruction/production can only resolve the \textit{net} effect of loss and resupply to Venus's atmosphere: species undergoing net atmospheric destruction may also be part of geological cycles that contribute both further sinks (e.g., by surface interaction) and sources (e.g., by volcanism, discussed later). Consequently, the photochemically-derived destruction rates provide a lower bound on the resupplying gas rates into the atmosphere to sustain steady-state: any additional geological sinks would also need to be balanced by resupply.  Most importantly for whether Venus's interior is wet, we need to consider whether there are additional geological sinks of H that would mean the required H flux back to the atmosphere is greater than we calculate (and therefore the \ce{OCS}/\ce{H2O} and \ce{CO2}/\ce{H2O} ratios of any restorative gas flux are lower than photochemical sinks alone imply). We consider how any such sinks may affect the photochemically derived fluxes below.

\subsection*{Sinks of the Venusian atmosphere}\label{sec:sinks}
Data from Venus's surface and sub-cloud atmosphere indicate ongoing chemical alteration of rocks and minerals by atmospheric gases, acting as an additional sink for some atmospheric species\cite{Filiberto}. Our understanding of chemical weathering of Venus's surface is derived from mineral compositions obtained through flybys, orbiters, and landers (e.g., refs \citenum{oyama1980pioneer, surkov1986venus, 2008Marcq}), and the predicted stability of minerals on the surface of Venus (e.g., refs \citenum{2017Radoman, berger2019experimental, teffeteller2022experimental}). The kinetics of these reactions are key for understanding their ability to shape the Venusian atmosphere. However, our knowledge of reaction rates is limited by the uncertain composition of the deep atmosphere and surface mineral compositions, and the need for experiments under Venusian conditions. Nonetheless, we can estimate the effect surface processes have on atmospheric sources and sinks of \ce{OCS}, \ce{CO2}, and \ce{H2O}, by reviewing the viability of candidate weathering reactions for each species in turn.

\ce{CO2}: Gas species in the near-surface atmosphere could oxidise exposed igneous rocks. Surface mineral oxidation could occur through reaction with atmospheric species \ce{CO2}, \ce{H2O} or \ce{SO2}\cite{Zolotov2019}. However, given the surface temperature, pressure and composition of the Venusian lower atmosphere\cite{FEGLEY1997416}, the equilibria of each of the redox reactions \ce{CO2 <=> CO + 1/2 O2}, \ce{SO2 <=> 1/2 S2 + O2}, and \ce{H2O <=> H2 + 1/2 O2}\cite{1977Ohmoto} are favoured to run in the direction of reducing the surface. The surface is thus unlikely to act as a sink for these oxidising agents. Were oxidation to occur, then it would be a competitive reaction between these three species.  Of these, \ce{CO2} is the prime candidate as it accounts for $\sim$96\% of the surface gas-phase composition, in contrast to \ce{SO2} at 150\,ppm and \ce{H2O} at 30\,ppm\cite{2008Marcq, pollack1993near}.  CO$_2$ removal could thus occur by oxidation of pyroxene ((Ca,Mg,Fe)SiO$_\text{3}$) into magnetite (Fe$_3$O$_4$), and magnetite into hematite (\ce{Fe2O3}) \cite{Zolotov2019}. Surface silicates on Venus have also been proposed to consume atmospheric \ce{CO2} through carbonitisation into carbonates. However, it is unclear whether this process is operating given the stability of carbonates at Venus's surface temperature\cite{TREIMAN2012534}. If carbonate–silicate equilibria are stable at Venus's conditions, they could serve as an additional \ce{CO2} removal mechanism. Importantly, this analysis implies that if surface oxidation or carbonitisation are occurring, it is most likely to require an \emph{upwards} revision of the \ce{CO2}/\ce{H2O} in the restorative gas --- i.e., towards a drier gas. 

\ce{OCS}: Carbonyl sulfide could also be subject to weathering sinks. It could react with the FeO in silicates, hematite, siderite (\ce{FeCO3}), magnetite, or pyrrhotite (\ce{Fe_{0.875}S}) to produce \ce{FeS2}\cite{Zolotov2019}. Though the reaction kinetics are uncertain, if operating these \ce{OCS} destruction fluxes would increase the \ce{OCS}/\ce{H2O} ratio of the overall gas removed from the atmosphere.

\ce{H2O}: Water would be lost from the atmosphere if it were involved in forming hydrous minerals.  However, there are no hydrous or H-bearing minerals stable at Venus's surface conditions (Supplementary Fig. 5). Consequently, there are no mechanisms to capture and retain the H from \ce{H2O} in a mineral form on the surface. Oxidation reactions with \ce{H2O} produce \ce{H2} as a by-product, such as \ce{3FeO(s) + H2O(g) -> Fe3O4(s) + H2(g)}  \cite{Zolotov2019}. The \ce{H2} can consequently serve as a reducing agent, as in \ce{H2(g) + CO2(g) -> CO(g) + H2O(g)}, effectively replenishing the reacted \ce{H2O} in the atmosphere. Yet, the present-day Venusian atmosphere indicates a surplus of \ce{H2} production (Supplementary Fig. 1), necessitating a means of \ce{H2} removal, not further production. This suggests crustal oxidation reactions are unlikely to create smaller (\ce{OCS},\ce{CO2})/\ce{H2O} ratios than estimated photochemically.

A second factor specific to the H budget of the atmosphere is escape to space. For a stable atmosphere, the flux of H escape needs to be balanced by higher H content in the restorative gas flux. This will require higher hydrogen content of the mantle than we have inferred from photochemical destruction alone.  The average rate of H-ion loss from Venus is 0.2--3.8\,$\times$\,10$^{25}$\,s$^{-1}$ \cite{RodriguezJ, HodgesH, BraceL, HuntenD, DonahueT, LAMMER20061445}.  This rate is not enough to overcome \ce{H2} production in the atmosphere, thereby still not requiring an \ce{H2} source via surface reaction of \ce{H2O}. The extreme scenario of H-loss rate where all escaped H is replenished by H from \ce{H2O} would require $\sim$0.2\% more \ce{H2O} in the replenishing gas. This remains inconsequential on the (\ce{OCS},\ce{CO2})/\ce{H2O} ratios, reinforcing the implication of a dry replenishing gas, and thus a dry Venusian interior.

\subsection*{Sources to the Venusian atmosphere}\label{sec:atmo_sources}

To replenish the gases undergoing atmospheric destruction, we explore a range of possible resupply mechanisms: only by ruling in volcanism and ruling out other mechanisms are we left with a constraint on the Venusian interior composition.

One possibility is that exogenic sources of material resupply \ce{OCS}, \ce{CO2} and \ce{H2O} to the atmosphere. We test this assuming a highly favourable case for this scenario: exogeneous delivery equivalent to the terrestrial flux of $\sim$7$\times$10$^{10}$\,g\,yr$^{-1}$ of pure carbonaceous chondrite material\cite{peucker1996accretion}.  This would deliver $\sim$1$\times$10$^{8}$\,moles\,yr$^{-1}$ of sulfur, $\sim$1$\times$10$^{8}$\,moles\,yr$^{-1}$ of carbon and $\sim$1$\times$10$^{9}$\,moles\,yr$^{-1}$ of hydrogen\cite{EBEL2000339}. If we again make the very favourable assumption that every mole of S is supplied as OCS, every mole of C as a mix of \ce{CO2} and \ce{OCS}, and every mole of H as \ce{H2O}, we can calculate the maximum impact of this delivery for the composition of any endogeneous atmospheric resupply.  For S the exogenic flux is four orders of magnitude smaller than the OCS destruction rate, for C four orders of magnitude smaller than the OCS and \ce{CO2} destruction rates, and for H three orders of magnitude smaller than the \ce{H2O} rate. Exogenous delivery is therefore clearly inadequate to counterbalance the destruction of sulfur, carbon, and hydrogen in the atmosphere.

On Earth, metamorphic degassing of buried sediments resupplies (predominantly) C, S, and H to the atmosphere.  The importance of this process on Venus is doubtful. For metamorphic devolatilisation to occur, two crucial conditions must be met: 1) mineral carriers of C, S, or H must first be stable at surface conditions in order to enter the crust, and 2) they must subsequently become unstable as they are buried to higher pressure and temperature environments\cite{kerrick2001metamorphic}.  The devolatilisation of H is implausible due the lack of stable H-bearing minerals at the surface (Supplementary Fig. 5). Similarly, C-bearing minerals like calcium carbonates are unlikely to be stable at Venusian surface conditions\cite{Zolotov2019}, so crustal devolatilisation is unlikely to be a source of gaseous H or C species.  

The story for S is more nuanced.  Magnetite–pyrite (\ce{Fe3O4}-\ce{FeS2}) and/or hematite (\ce{Fe\textsubscript{2}O\textsubscript{3}})–magnetite assemblages can be stable at the conditions of Venus’ plains, with pyrite also potentially stable in the highlands\cite{Zolotov2019}. It is not clear however, the extent to which sulfidation of the crust is active; this in part reflects uncertainties on the thermodynamics and surface composition. However, even if S and O can be present in minerals that are found to be stable at surface conditions, and their breakdown were to release OCS, the process of burial necessitates active subduction which is notably absent on modern Venus\cite{2022Junxing}. 

Alternative tectonic regimes for Venus have been proposed, such as the ``plutonic-squishy lid"\cite{2020Lour} model driven by intrusive magmatism. While this regime might involve short-lived, sporadic episodes of ductile plate sinking, it is thought to lack the large-scale downwellings characteristic of plate tectonics\cite{2020Lour}. In the absence of planetary-scale tectonic burial, the only remaining mechanism would be volcanic burial.  However, in that case the volcanic gas released from the magmas is likely to be as important, if not more so, than resultant metamorphic degassing. We therefore conclude that the gas to balance photochemical destruction, if the atmosphere is in steady state, must come from volcanism.

\section*{Volcanic gas chemistry on Venus}\label{sec:volcanic}

There is good evidence that Venus continues to be volcanically active.  Venus's surface is widely recognised as having been shaped by volcanism\cite{stofan1992, Barsukov1992, Surkovvenera}.  Evidence for ongoing volcanism comes from a number of sources: an observed lava flow at a volcanic vent\cite{Herrick2023}, young lava flows based on surface ageing from emissivity measurements\cite{Filiberto}, and volcanic degassing is necessary to maintain the global sulfuric acid (\ce{H2SO4}) clouds\cite{fegley1989estimation}.  The key question we address here is what the average composition of this volcanism is, and in particular, its water content. The key ratios of species being photochemically destroyed, OCS/\ce{H2O} and \ce{CO2}/\ce{H2O}, imply a very dry volcanic gas, and by extension, a dry magma. Interpreting the magma composition as a reliable proxy for the globally-averaged supplying mantle, this suggests a correspondingly dry interior. We test this hypothesis by modelling volcanic degassing speciation of a dry terrestrial basalt at Venus's surface conditions (\nameref{sec:methods}), to determine if such a dry gas can be consistent with Earth-like magmas.

\begin{figure*}[htbp]
	\includegraphics[width=\columnwidth]{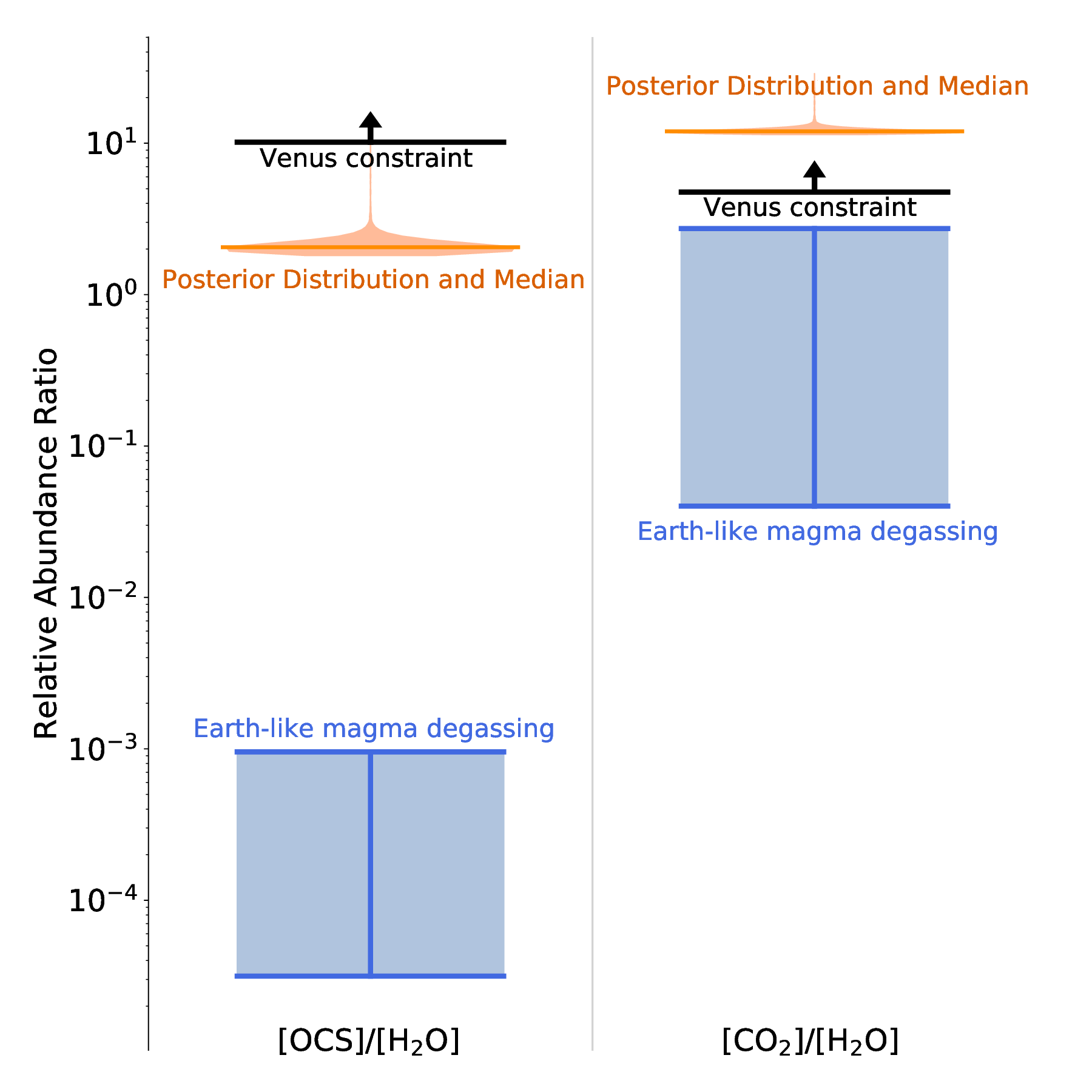}
 \caption{ \textbf{Constrained Venusian volcanic gas compared with Earth-like magma degassing.} Ratio of \ce{OCS} and \ce{CO2} abundances relative to \ce{H2O} abundance (\ce{OCS}/\ce{H2O} and \ce{CO2}/\ce{H2O} respectively) in Venus's restorative volcanic gas, constrained directly from atmospheric destruction fluxes of \ce{OCS}, \ce{H2O} and \ce{H2O} (black lines). The relative abundances provide a conservative lower limit for the volcanic gas speciation due to the potential surface weathering sinks affecting \ce{OCS} and \ce{CO2}, necessitating a higher restorative degassing flux to balance them, while \ce{H2O} is unlikely to be affected by surface sinks from weathering. The posterior distribution obtained from the Markov Chain Monte Carlo sampler, considering variations in degassing pressure and temperature (represented by orange violins), with the horizontal orange line denoting the median value. This median value signifies the equilibrium gas composition that best aligns with the atmospheric speciation constraints. The blue bars indicate the range of \ce{OCS}/\ce{H2O} and \ce{CO2}/\ce{H2O}, when degassing a terrestrial basalt under different Venusian pressure conditions, ranging from the deep interior to the surface pressure.}
 \label{fig:violin}
\end{figure*}

\begin{figure*}[htbp]
	\includegraphics[width=\columnwidth]{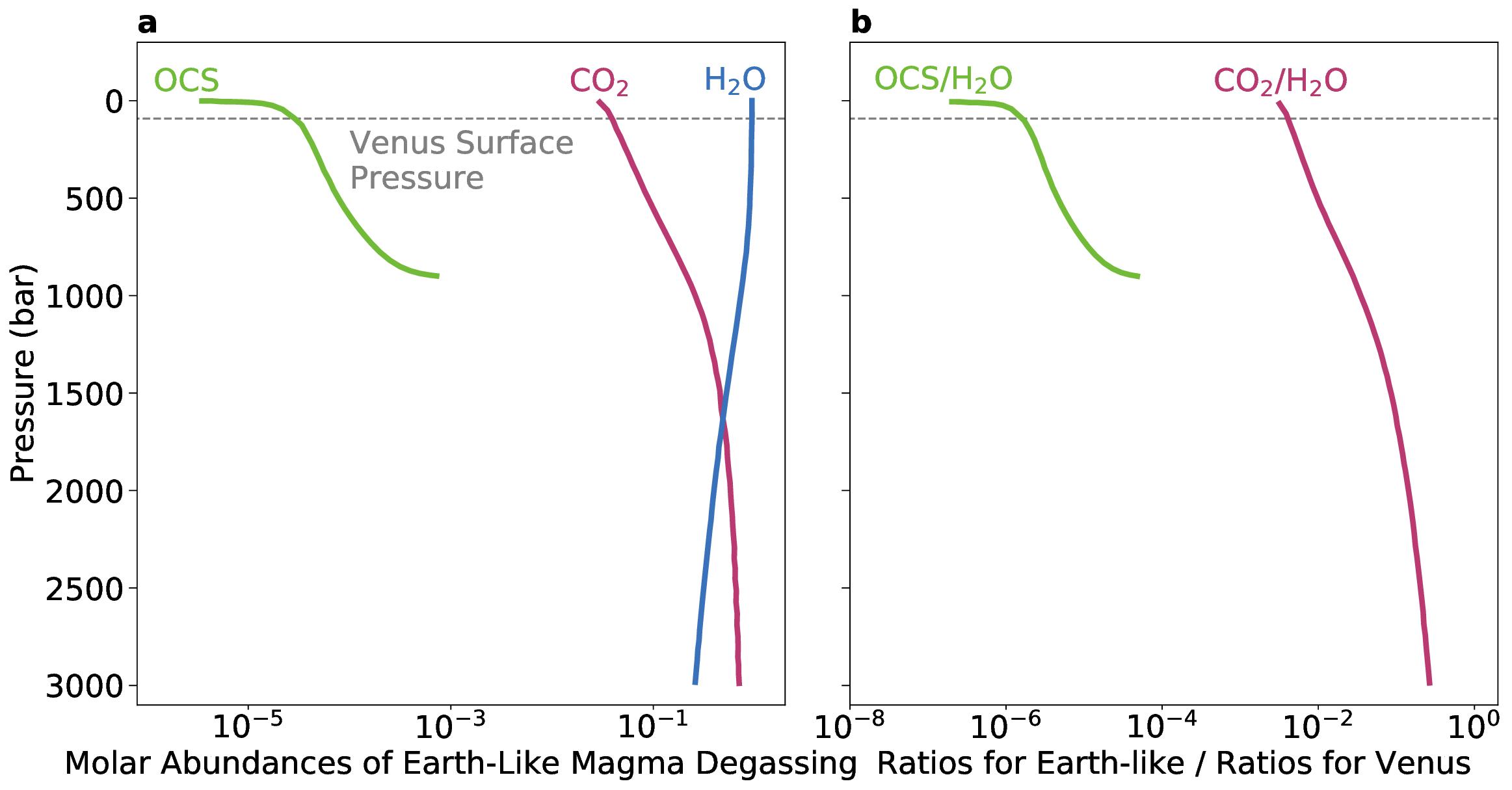}
 \caption{ \textbf{Volcanic gas composition comparison across pressure. a,} Variation in MORB degassing speciation under different pressure conditions, plotted against the volcanic gas molar abundances of \ce{OCS}, and \ce{CO2}, relative to \ce{H2O}. \textbf{b,} Ratio of relative molar abundances \ce{OCS}/\ce{H2O} and \ce{CO2}/\ce{H2O} degassed from Earth's MORB, to \ce{OCS}/\ce{H2O} and \ce{CO2}/\ce{H2O} from Venus' volcanic gas, as discussed in Section \nameref{sec:volcanic}. Venus' surface pressure plotted for reference (black dashed line).}
 \label{fig:evo_misfit}
\end{figure*}

Figure \ref{fig:violin} (blue bars) shows the speciation of the volcanic gas degassed by the dry terrestrial basalt at varying Venusian pressures. In comparison to the Earth-like volcanic gas, characterised by a 96\% molar water content degassed at Venus's surface pressure (or a minimum of 25\% when degassed at a higher pressure of 3000\,bar) we find a distinct contrast when assessing Venus's volcanic gases from the atmospheric composition constraints, revealing a maximum water molar content of 6\%. The \ce{CO2}/\ce{H2O} abundance offset between Venus's volcanic gas and Earth's MORB volcanic gas decreases with increasing degassing pressure (Fig \ref{fig:evo_misfit}), though it is critical to note that the Venusian values represent lower limits as we have not accounted for potential weathering sinks of OCS and \ce{CO2} (see above). The \ce{OCS}/\ce{H2O} shows the largest discrepancy with terrestrial basalt, being at least 4 orders of magnitude too low in the terrestrial magma, so 4 orders of magnitude more \ce{H2O}. This significant offset in abundances in the volcanic gases indicates that the Venusian magma lacks \ce{H2O}; instead being sulfur and carbon rich. 

If the restorative gas is supplied through volcanism, it is likely degassed at high temperatures and pressures, and thus in local thermodynamic equilibrium. By taking a system with composition defined by OCS, \ce{CO2} and \ce{H2O} in their inferred relative proportions, we vary pressure and temperature to examine whether these species can be found in thermochemical equilibrium in their inferred abundances (\nameref{sec:methods}). We find that the inferred (\ce{OCS},\ce{CO2})/\ce{H2O} ratios are matched within an order of magnitude by a temperature of $\sim$1890\,K and a pressure of $\sim$60\,bar (Fig. \ref{fig:violin}). This alignment suggests that the observed flux ratios originate from a high-temperature process, consistent with volcanism.

\section*{Implications for the climate histories of Venus-like planets} \label{sec:history}

A dry Venusian interior is not consistent with Venus having had surface oceans, or by extension a conventionally habitable climate. The most plausible way for both the atmosphere and interior to have efficiently lost the originally-accreted H, is if the water reservoir existed as a steam atmosphere on the timescale of tens of millions of years above a magma ocean, leaving a long-lived dry surface\cite{hamano2013emergence}. 

D/H ratios have been a historically important constraint on the water history of Venus\cite{donahue1997venus}.  The climate history we infer here puts this central observation in new light: long term desiccation of the planet is most consistent with a steady state source of continuous outgassing to the atmosphere \cite{GrinspoonDH}. The D/H ratio of the water being outgassed would have to be substantially higher than that of terrestrial water.  This would be consistent with degassing from a highly fractionated mantle source, as would be the case if the mantle inherited the fractionation from the early coupled atmosphere-magma ocean loss episode\cite{hamano2013emergence}, when the planet first dried out.  

These results indicate that Venus likely never experienced conditions conducive to ocean condensation. A result in agreement with recent modelling work indicating that a Venus born with a hot steam atmosphere would never have received a low enough insolation to condense its water reservoir\cite{turbet2023water}. Consequently, Venus-like exoplanets or planets within the Venus-zone\cite{Kane2019} that JWST will be characterising, are unlikely to be cool enough to condense liquid water if they formed in situ. This makes these planets improbable candidates for `liquid-water' habitable conditions.

\section*{Methods}\label{sec:methods}

\subsection*{Modelling the Venusian Atmosphere} \label{sec:atmosphere}

To model the atmosphere, a 1D Lagrangian High Energy photochemistry-diffusion model\cite{Rimmer2016} is used, called ARGO, as a numerical solver for STAND2022, an atmospheric chemical network describing gas phase photo-chemical kinetics\cite{Rimmer2016, rimmer2021hydroxide}. STAND includes over 6600 thermochemical and photochemical reactions involving 480 species composed of C-O-H-S-N-Cl and a handful of other elements. The chemical kinetics code has been validated to within an order of magnitude of most observations of atmospheric species abundances for Venus.
 
The reactions from STAND are solved by ARGO, as a set of time-dependent, coupled, non-linear differential equations:
\begin{equation}\label{eqn:partial}
\frac{\partial n_X}{\partial t} = P_X - L_Xn_X - \frac{\partial \Phi_{diff,X}}{\partial z}
\end{equation}
at height $z$ (cm) from the surface, where $n_X$ (cm$^{-3}$) is the number density of a species $X$, $P_{X}$ (cm$^{-3}$ s$^{-1}$) is the rate of production of species $X$, $L_{X}$ (s$^{-1}$) is the rate constant for destruction of species $X$, and $\Phi_{diff,X}$ (cm$^{-2}$\,s$^{-1}$) is the vertical flux of species as a function of eddy- and molecular-diffusion processes. $P_{X}$ and $L_{X}$ are determined from the rate constants of the different reactions within STAND, and the relevant species abundances. $\frac{\partial n_X}{\partial t}$ represents the rate of change of number density of species $X$ at each height $z$ in the atmosphere, as a function of the species' production, destruction and the change of the species' vertical flux with respect to height. 

These chemical reactions solved by ARGO determine the chemical production and loss rates of each species at each altitude level in the atmosphere. ARGO traces the path of a gas parcel as it moves from the surface to the top of the atmosphere and back down. It starts with initial chemical composition data at the base of the atmosphere. At each altitude level during the upward journey, ARGO solves Eq. \ref{eqn:partial} for all atmospheric species using pressure, temperature, and species abundances. The time interval for solving is determined by the eddy-diffusion profile, which accounts for vertical transport. The pressure-temperature and eddy-diffusion profiles for modern Venus are adopted from previous photochemical-kinetics models of the lower and middle atmosphere. At the top of the atmosphere, ARGO incorporates the incident stellar spectrum and includes photochemical reactions driven by the stellar flux during the downward journey. The process is iterated until a convergence criterion is met, leading to a comprehensive global solution.

\subsection*{Atmospheric flux and timescale measurements} \label{sec:fluxes}

The net photochemical and thermochemical production or destruction of $X$ in the atmosphere, $\Phi_{atmo,X}$ (cm$^{-2}$ s$^{-1}$), is used to determine the chemistry of the restorative gas flux in the atmosphere (Sec. \nameref{sec:atmo_losses}). $\Phi_{atmo,X}$ is calculated for each species $X$ by summing over the atmospheric chemical production and destruction of $X$ at each altitude step at height $z$, as determined by ARGO, where
\begin{equation}\label{eqn:flux_eqn}
\Phi_{atmo,X}(z) = \Phi_{photochem,X}(z) + \Phi_{thermochem,X}(z) + \Phi_{diffusion,X}(z)
\end{equation}
The change in number density is 
\begin{equation}
\frac{d n_X(0,t)}{dt} = \frac{d \Phi_{X}(t) }{H_{sc} }
\end{equation}
at surface pressure, where $\Phi_{X}$ (cm$^2$\,s$^{-1}$) is the total species flux in the atmosphere, and $H_{sc}$.is the atmospheric scale height. Assuming steady state, each species has a fixed number density, $\frac{d n_X(0, t)}{dt} = 0$, so $\Delta \Phi_{X}$ is zero for all $X$. Within our atmospheric model, the bottom of the atmosphere is incorporated as fixed initial surface mixing ratios. These are chosen such that the model fits observational measurements of species abundances further up in Venus's atmosphere\cite{rimmer2021hydroxide}.

Atmospheric processes alone are not sufficient to describe all processes affecting atmospheric gas fluxes $\Phi_{X}$, as the model does not include interactions with the surface or atmospheric escape. The net atmospheric flux is
\begin{equation}\label{eqn:total_flux}
\Delta\Phi_X = \Phi_{deg,X} - \Phi_{esc,X}
- \Phi_{sink,X}+ \Phi_{atmo,X} = 0
\end{equation}
where $\Phi_{deg,X}(t)$ (cm$^{-2}$ s$^{-1}$) is the surface volcanic degassing of $X$, $\Phi_{esc,X}$ (cm$^{-2}$ s$^{-1}$) is the escape flux, $\Phi_{sink,X}$ (cm$^{-2}$ s$^{-1}$) is the deposition flux due to interactions with the surface, and $\Phi_{atmo,X}$ (cm$^{-2}$ s$^{-1}$) is the net photochemical and thermochemical production or destruction of $X$ in the atmosphere, as calculated by our model. 

This is used to constrain $\Phi_{deg,X}$ to gain insight into the volcanic degassing speciation on Venus.  There can be still redundancy, however, in solving for Eq. \ref{eqn:total_flux} as the weathering reaction rates $\Phi_{sink, X}$ for the Venusian surface conditions are still mainly unknown, as addressed in Section \nameref{sec:sinks}.

The $\Phi_{atmo,X}$ fluxes (Fig. \ref{fig:flux_time}, top subfigure) integrate photochemical, thermochemical, and diffusive processes within the atmosphere
\begin{equation}
\begin{split}
 \Phi_{atmo, X} = \int_{0}^{\infty} \left[ P_X(z,t)- L_X(z,t)n_X(z,t) - \frac{\partial \Phi_{diff,X}}{{\partial z}}\right] dz. \\
\end{split}
\end{equation}
To extract $\Phi_{atmo,X}$ from the model, we track the history of the air parcel throughout the cycle by calculating the change in molecular abundance in the parcel at the start of the gas parcel cycle compared to the end ($\Delta$n) when convergence is reached as the parcel reaches the surface again.
\begin{equation} \label{eqn:full_flux_eqn} 
\begin{split}
 \Phi_{atmo, X} = \frac{2 \Delta n K_{ZZ}}{H_{sc} } \\
\end{split}
\end{equation}
The coefficients used to construct P$_X$ and L$_X$ are provided by the chemical network, STAND. The velocity of the gas parcel in the model is determined through $\frac{2K_{ZZ}}{H_{sc}}$, where K$_{ZZ}$ is the eddy diffusion at the surface (cm$^{2}$ s$^{-1}$).

The e-folding time of atmospheric species is determined from their expected lifetime in the atmosphere (Fig. \ref{fig:flux_time}, bottom subfigure), defined as the species column density divided by $\Phi_{atmo,X}$. 

The most dominant reactions for each species are determined through comparing the column rates $R_{chem}$ (cm$^{-3}$ s$^{-1}$) of individual reactions (Plotted in Supplementary Figs. 3--5). At the last time-instance that the parcel spends at each height $z$,  
\begin{equation}\label{eqn:rate} 
R_{chem} =  k(z) \prod_{i = 1}^{N} n_{X_{i}}\\
\end{equation}
where $k(z)$ (cm$^{3N-3}$ s$^{-1}$) is the reaction rate constant extracted from STAND (right before the parcel moves, see Supplementary Text 2), and N is the total number of reactants.

\subsection*{Modelling volcanic degassing of a dry terrestrial basalt}\label{sec:method_volc}

The speciation of a degassing a dry terrestrial basalt is simulated using a thermodynamic magma degassing model, EVo\cite{liggins2020can, Liggins2022, Liggins2022b}. A Mid-Ocean Ridge Basalt (MORB) serves as a valuable reference point for the ambient mantle of a plate-tectonically active planet, far away from direct inputs of wet material at subduction zones. EVo iteratively calculates melting and subsequent volatile outgassing of a gas in equilibrium with its parent a mid-ocean ridge basaltic magma. The melting model within EVo uses the batch melting equation (see ref. \citenum{liggins2020can}) to partition volatiles from the bulk mantle into the melt phase during each time step. 

This model calculates the speciation and volume of a C-O-H-S gas phase, including our key species \ce{OCS}, \ce{CO2} and \ce{H2O}, in equilibrium with a silicate melt at a given pressure, temperature and magma \ce{fO2}, considering both homogeneous gas-phase equilibria and heterogeneous gas-melt equilibria considered in the form of solubility laws suitable for a melt with a basaltic composition (detailed across refs. \citenum{liggins2020can, Liggins2022, Liggins2022b}). The only exception is for \ce{OCS}, where its abundance is calculated at an additional gas-phase thermodynamic equilibrium step, as there is a lack of solubility data in magmas to be considered for a melt-gas equilibrium.

EVo was initialised with volatile contents characteristic of a MORB: 0.3 wt\% \ce{H2O} and 0.01 wt\% total sulfur (as a sum of dissolved \ce{SO2} and \ce{H2S}). The f\ce{O2} (oxygen fugacity) was set to FMQ-0.5, which aligns well with the range observed in erupting MORBs on Earth (FMQ-0.41$\pm$0.43\cite{bezos2005fe3}). The model simulated degassing at a fixed temperature of 1400\,K (representing the average temperature of degassing magma) and a pressure range spanning beyond potentially relevant Venusian conditions – from deep within the interior at 3000\,bar, to the surface pressure on Earth 1\,bar.

\subsection*{Surface-Atmosphere thermo-chemical equilibrium} \label{sec:ggchem}

We examine the stability of H-bearing minerals at the surface using GG$\sf{_{CHEM}}$ \cite{woitke2018equilibrium}, by considering a mixture of gas and condensed species at 92\,bars, with surface elemental abundances estimated from Vega 2 surface oxide ratios and the surface gas abundances in our atmospheric model (as in ref. \citenum{rimmer2021hydroxide}). Supplementary Figure 5 provides insight into the stable H-bearing condensates across a temperature range, encompassing both the cooling process from molten magma to Venus's surface temperature and extending to lower temperatures where solid phases become more stable. At Venus's surface conditions, H exists solely in its gaseous form (Supplementary Fig. 5, represented by the blue line). Water in the magma is unstable under Venus's surface conditions and preferentially exsolves (Fig. \ref{fig:evo_misfit}).  Even if water were to remain in the magma after degassing, it would eventually diffuse out, potentially becoming a long-term source of water in the atmosphere, as the absence of stable hydrous or H-bearing minerals at means that the surface cannot retain hydrogen. This gaseous release into the atmosphere would still be included in the resupplying \ce{H2O} flux we constrained atmospherically. Therefore, based on the understanding that we cannot reproduce the atmospherically-constrained volcanic water content through magma degassing, and that the water cannot be retained in the surface within solids or minerals, we conclude that the surfacing magma itself must be dry.

\subsection*{Bayesian sampling for volcanic gas speciation at thermochemical equilibrium} \label{sec:mcmc}

We investigate the closest-fit composition of the atmospherically-inferred volcanic gas to a gas in thermochemical equilibrium using Fastchem\cite{fastchem}. Using the elemental abundances inferred from the proportions of OCS, \ce{CO2}, and \ce{H2O}, we varied pressure and temperature to determine whether these species can be found in thermochemical equilibrium in their inferred abundances. We use the emcee affine-invariant ensemble sampler for Markov Chain Monte Carlo sampling (MCMC)\cite{emcee}, to sample volcanic gas pressure and temperature priors and determine the best fit for our atmospheric constraints. ${\chi}^2$ is used as a measure of the goodness of fit between the gas speciation inferred from our atmospheric model, to the equilibrium gas speciation produced by Fastchem. 

The priors are defined by the range of pressures over which a magma can be degassed, from deeper within the planetary interior at 3000\,bar, to pressures higher up in the atmosphere at 60\,bar. Similarly for the temperature range, ranging from 1900\,K to a cooler 1000\,K as the magma cools down. Although the most-likely priors are at the edge of the prior distribution, suggesting that further extremes might be a better fit, extending the distribution would not correspond to feasible degassing conditions.  The complete resulting gas chemistry of the most probable gas is primarily composed of \ce{CO}, \ce{CO2}, \ce{S2}, \ce{OCS}, \ce{H2S} and \ce{H2O}. This suggests that The constraining speciation ratios used in the ${\chi}^2$ measure of fit would differ following the incorporation of weathering sinks of the key species (as shown by the upward arrows in Fig. \ref{fig:violin}, indicating that weathering would shift the fluxes to drier ratios).

\section*{Data Availability}
No new observational data were generated for this work. The data used to produce the figures related to Venus's atmospheric chemistry can be found across refs. \citenum{Rimmer2016, rimmer2019hydrogen, rimmer2021hydroxide, hobbs2021sulfur}. The data needed to evaluate the findings have been deposited in the Harvard online database under accession code \url{https://doi.org/10.7910/DVN/19HF6Q}.

\section*{Code Availability}
The methods underlying ARGO and STAND are comprehensively documented in a collection of publicly available papers: \citenum{Rimmer2016, rimmer2019hydrogen, rimmer2021hydroxide, hobbs2021sulfur}. The ARGO software with the atmospheric flux calculation extension may be made available upon request. The EVo version used here is publicly available at \url{https://github.com/pipliggins/EVo} and is documented across refs. \citenum{liggins2020can, Liggins2022}. The FastChem 2.0 model \cite{fastchem} is freely available at \url{https://github.com/exoclime/FastChem}, where updates are hosted. GG$\sf{_{CHEM}}$ \cite{woitke2018equilibrium} is also freely available and updated at \url{https://github.com/pw31/GGchem}. 

\section*{Acknowledgements}
T.C. thanks the Science and Technology Facilities Council (STFC) for the PhD studentship (grant reference ST/X508299/1). The authors thank Philippa Liggins, Craig Walton, Skyla White, Claire Marie Guimond, Avishek Rudra, and Sean Jordan for helpful discussions. 

\section*{Corresponding authors}
Correspondence to \href{mailto:tc496@cam.ac.uk}{Tereza Constantinou}.

\section*{Author Contributions Statement}
T.C., O.S. and P.B.R. built the conceptualisation of study. T.C. performed the modelling and analysis. O.S. and P.B.R. supervised the project. All authors contributed to the writing of this paper. P.B.R. focused their contribution to the atmospheric chemistry aspects of the paper.

\section*{Competing Interests Statement}
The authors declare no competing interests.


\begin{thebibliography}{10}
\expandafter\ifx\csname url\endcsname\relax
  \def\url#1{\burl{#1}}\fi
\expandafter\ifx\csname urlprefix\endcsname\relax\def\urlprefix{URL }\fi
\providecommand{\bibinfo}[2]{#2}
\providecommand{\eprint}[2][]{\url{#2}}


\bibitem{carr1987water}
\bibinfo{author}{Carr, M.~H.}
\newblock \bibinfo{title}{Water on mars}.
\newblock \emph{\bibinfo{journal}{Nature}} \textbf{\bibinfo{volume}{326}}, \bibinfo{pages}{30--35} (\bibinfo{year}{1987}).

\bibitem{Strom1994}
\bibinfo{author}{Strom, R.~G.}, \bibinfo{author}{Schaber, G.~G.} \& \bibinfo{author}{Dawson, D.~D.}
\newblock \bibinfo{title}{The global resurfacing of {Venus}}.
\newblock \emph{\bibinfo{journal}{J. Geophys. Res.: Planets}} \textbf{\bibinfo{volume}{99}}, \bibinfo{pages}{10899--10926} (\bibinfo{year}{1994}).

\bibitem{donhue}
\bibinfo{author}{Donahue, T.~M.}, \bibinfo{author}{Hoffman, J.~H.}, \bibinfo{author}{Hodges, R.~R.} \& \bibinfo{author}{Watson, A.~J.}
\newblock \bibinfo{title}{{Venus} was wet: A measurement of the ratio of deuterium to hydrogen}.
\newblock \emph{\bibinfo{journal}{Science}} \textbf{\bibinfo{volume}{216}}, \bibinfo{pages}{630--633} (\bibinfo{year}{1982}).

\bibitem{raymond2017}
\bibinfo{author}{Raymond, S.~N.}, \bibinfo{author}{Quinn, T.} \& \bibinfo{author}{Lunine, J.~I.}
\newblock \bibinfo{title}{High-resolution simulations of the final assembly of {Earth}-like planets. 2. {Water} delivery and planetary habitability}.
\newblock \emph{\bibinfo{journal}{Astrobiology}} \textbf{\bibinfo{volume}{7}}, \bibinfo{pages}{66--84} (\bibinfo{year}{2007}).
\newblock \bibinfo{note}{PMID: 17407404}.

\bibitem{donahue1997venus}
\bibinfo{author}{Donahue, T.~M.} \& \bibinfo{author}{Russell, C.~T.}
\newblock \bibinfo{title}{ in \textit{The {Venus} atmosphere and ionosphere and their interaction with the solar wind: {An} overview}} (eds \bibinfo{editor}{Baugher, S.~W.}, \bibinfo{editor}{Hunten, D.~M.} \& \bibinfo{editor}{Phillips, R.~J.}) \emph{\bibinfo{booktitle}{Venus II Geology, Geophysics, Atmosphere, and Solar Wind Environment}} \bibinfo{pages}{3--31} (\bibinfo{publisher}{Univ. of Ariz. Press, Tucson}, \bibinfo{year}{1997}).

\bibitem{Warren}
\bibinfo{author}{Warren, A.~O.} \& \bibinfo{author}{Kite, E.~S.}
\newblock \bibinfo{title}{Narrow range of early habitable {Venus} scenarios permitted by modeling of oxygen loss and radiogenic argon degassing}.
\newblock \emph{\bibinfo{journal}{Proc. Natl Acad. Sci. USA}} \textbf{\bibinfo{volume}{120}}, \bibinfo{pages}{e2209751120} (\bibinfo{year}{2023}).

\bibitem{grinspoon1988cometary}
\bibinfo{author}{Grinspoon, D.~H.} \& \bibinfo{author}{Lewis, J.~S.}
\newblock \bibinfo{title}{Cometary water on {V}enus: Implications of stochastic impacts}.
\newblock \emph{\bibinfo{journal}{Icarus}} \textbf{\bibinfo{volume}{74}}, \bibinfo{pages}{21--35} (\bibinfo{year}{1988}).

\bibitem{1988Kasting}
\bibinfo{author}{{Kasting}, J.~F.}
\newblock \bibinfo{title}{Runaway and moist greenhouse atmospheres and the evolution of {Earth} and {Venus}}.
\newblock \emph{\bibinfo{journal}{Icarus}} \textbf{\bibinfo{volume}{74}}, \bibinfo{pages}{472--494} (\bibinfo{year}{1988}).

\bibitem{GrinspoonBullock2007}
\bibinfo{author}{Grinspoon, D.~H.} \& \bibinfo{author}{Bullock, M.~A.}
\newblock \bibinfo{title}{Astrobiology and {Venus} exploration} 
\newblock \emph{\bibinfo{journal}{Geophys. Monogr.-Ame. Geophys. Union}} (\bibinfo{year}{2007}).

\bibitem{Way2016}
\bibinfo{author}{Way, M.~J.} \emph{et~al.}
\newblock \bibinfo{title}{Was {V}enus the first habitable world of our solar system?}
\newblock \emph{\bibinfo{journal}{Geophys. Res. Lett.}}\textbf{\bibinfo{volume}{43}}, \bibinfo{pages}{8376--8383} (\bibinfo{year}{2016}).

\bibitem{Way2020}
\bibinfo{author}{Way, M.~J.} \& \bibinfo{author}{Genio, A. D.~D.}
\newblock \bibinfo{title}{{Venusian} habitable climate scenarios: Modeling {Venus} through time and applications to slowly rotating {Venus}-like exoplanets}.
\newblock \emph{\bibinfo{journal}{J. Geophys. Res.: Planets}} \textbf{\bibinfo{volume}{125}} (\bibinfo{year}{2020}).

\bibitem{BULLOCK200119}
\bibinfo{author}{Bullock, M.~A.} \& \bibinfo{author}{Grinspoon, D.~H.}
\newblock \bibinfo{title}{The recent evolution of climate on {V}enus}.
\newblock \emph{\bibinfo{journal}{Icarus}} \textbf{\bibinfo{volume}{150}}, \bibinfo{pages}{19--37} (\bibinfo{year}{2001}).

\bibitem{hamano2013emergence}
\bibinfo{author}{Hamano, K.}, \bibinfo{author}{Abe, Y.} \& \bibinfo{author}{Genda, H.}
\newblock \bibinfo{title}{Emergence of two types of terrestrial planet on solidification of magma ocean}.
\newblock \emph{\bibinfo{journal}{Nature}} \textbf{\bibinfo{volume}{497}}, \bibinfo{pages}{607--610} (\bibinfo{year}{2013}).

\bibitem{GILLMANN2009503}
\bibinfo{author}{Gillmann, C.}, \bibinfo{author}{Chassefi\'ere, E.} \& \bibinfo{author}{Lognonn\'e, P.}
\newblock \bibinfo{title}{A consistent picture of early hydrodynamic escape of {Venus} atmosphere explaining present ne and ar isotopic ratios and low oxygen atmospheric content}.
\newblock \emph{\bibinfo{journal}{Earth Planet. Sci. Lett.}} \textbf{\bibinfo{volume}{286}}, \bibinfo{pages}{503--513} (\bibinfo{year}{2009}).

\bibitem{Turbet2021}
\bibinfo{author}{Turbet, M.} \emph{et~al.}
\newblock \bibinfo{title}{Day\-night cloud asymmetry prevents early oceans on {V}enus but not on {E}arth}.
\newblock \emph{\bibinfo{journal}{Nature}} \textbf{\bibinfo{volume}{598}}, \bibinfo{pages}{276--280} (\bibinfo{year}{2021}).

\bibitem{selsis2023cool}
\bibinfo{author}{Selsis, F.}, \bibinfo{author}{Leconte, J.}, \bibinfo{author}{Turbet, M.}, \bibinfo{author}{Chaverot, G.} \& \bibinfo{author}{Bolmont, {\'E}.}
\newblock \bibinfo{title}{A cool runaway greenhouse without surface magma ocean}.
\newblock \emph{\bibinfo{journal}{Nature}} \textbf{\bibinfo{volume}{620}}, \bibinfo{pages}{287--291} (\bibinfo{year}{2023}).

\bibitem{Gillmann2022}
\bibinfo{author}{Gillmann, C.} \emph{et~al.}
\newblock \bibinfo{title}{The long-term evolution of the atmosphere of {V}enus: processes and feedback mechanisms}.
\newblock \emph{\bibinfo{journal}{Space Sci. Rev.}}  (\bibinfo{year}{2022}).

\bibitem{gillmann2020dry}
\bibinfo{author}{{Gillmann}, C.} \emph{et~al.}
\newblock \bibinfo{title}{{Dry late accretion inferred from Venus's coupled atmosphere and internal evolution}}.
\newblock \emph{\bibinfo{journal}{Nat. Geosci.}} \textbf{\bibinfo{volume}{13}}, \bibinfo{pages}{265--269} (\bibinfo{year}{2020}).

\bibitem{dorn2021hidden}
\bibinfo{author}{Dorn, C.} \& \bibinfo{author}{Lichtenberg, T.}
\newblock \bibinfo{title}{Hidden water in magma ocean exoplanets}.
\newblock \emph{\bibinfo{journal}{Astrophys. J. Lett.}} \textbf{\bibinfo{volume}{922}}, \bibinfo{pages}{L4} (\bibinfo{year}{2021}).

\bibitem{bower2022retention}
\bibinfo{author}{Bower, D.~J.}, \bibinfo{author}{Hakim, K.}, \bibinfo{author}{Sossi, P.~A.} \& \bibinfo{author}{Sanan, P.}
\newblock \bibinfo{title}{Retention of water in terrestrial magma oceans and carbon-rich early atmospheres}.
\newblock \emph{\bibinfo{journal}{Planet. Sci. J.}} \textbf{\bibinfo{volume}{3}}, \bibinfo{pages}{93} (\bibinfo{year}{2022}).

\bibitem{Miyazaki}
\bibinfo{author}{Miyazaki, Y.} \& \bibinfo{author}{Korenaga, J.}
\newblock \bibinfo{title}{Inefficient water degassing inhibits ocean formation on rocky planets: An insight from self-consistent mantle degassing models}.
\newblock \emph{\bibinfo{journal}{Astrobiology}} \textbf{\bibinfo{volume}{22}}, \bibinfo{pages}{713--734} (\bibinfo{year}{2022}).

\bibitem{luo2024majority}
\bibinfo{author}{Luo, H.}, \bibinfo{author}{Dorn, C.} \& \bibinfo{author}{Deng, J.}
\newblock \bibinfo{title}{Majority of water hides deep in the interiors of exoplanets}.
\newblock \emph{\bibinfo{journal}{Preprint at https://arxiv.org/abs/2401.16394}}  (\bibinfo{year}{2024}).

\bibitem{armann2012simulating}
\bibinfo{author}{Armann, M.} \& \bibinfo{author}{Tackley, P.J.}
\newblock \bibinfo{title}{Simulating the thermochemical magmatic and tectonic evolution of Venus's mantle and lithosphere: Two-dimensional models}.
\newblock \emph{\bibinfo{journal}{J. Geophys. Res.: Planets}} 
\textbf{\bibinfo{volume}{117}}, (\bibinfo{year}{2012}).


\bibitem{GUIMOND2021106788}
\bibinfo{author}{Guimond, C.~M.}, \bibinfo{author}{Noack, L.}, \bibinfo{author}{Ortenzi, G.} \& \bibinfo{author}{Sohl, F.}
\newblock \bibinfo{title}{Low volcanic outgassing rates for a stagnant lid {A}rchean earth with graphite-saturated magmas}.
\newblock \emph{\bibinfo{journal}{Phys. Earth Planet. Inter.}} \textbf{\bibinfo{volume}{320}}, \bibinfo{pages}{106788} (\bibinfo{year}{2021}).

\bibitem{KrissansenTotton_2021}
\bibinfo{author}{Krissansen-Totton, J.}, \bibinfo{author}{Galloway, M.~L.}, \bibinfo{author}{Wogan, N.}, \bibinfo{author}{Dhaliwal, J.~K.} \& \bibinfo{author}{Fortney, J.~J.}
\newblock \bibinfo{title}{Waterworlds probably do not experience magmatic outgassing}.
\newblock \emph{\bibinfo{journal}{Astrophys. J.}} \textbf{\bibinfo{volume}{913}}, \bibinfo{pages}{107} (\bibinfo{year}{2021}).

\bibitem{kaula1999constraints}
\bibinfo{author}{Kaula, W.~M.}
\newblock \bibinfo{title}{Constraints on {V}enus evolution from radiogenic argon}.
\newblock \emph{\bibinfo{journal}{Icarus}} \textbf{\bibinfo{volume}{139}}, \bibinfo{pages}{32--39} (\bibinfo{year}{1999}).

\bibitem{o2015thermal}
\bibinfo{author}{O'Rourke, J.~G.} \& \bibinfo{author}{Korenaga, J.}
\newblock \bibinfo{title}{Thermal evolution of {Venus} with argon degassing}.
\newblock \emph{\bibinfo{journal}{Icarus}} \textbf{\bibinfo{volume}{260}}, \bibinfo{pages}{128--140} (\bibinfo{year}{2015}).

\bibitem{Aubaud2004}
\bibinfo{author}{Aubaud, C.}, \bibinfo{author}{Hauri, E.~H.} \& \bibinfo{author}{Hirschmann, M.~M.}
\newblock \bibinfo{title}{Hydrogen partition coefficients between nominally anhydrous minerals and basaltic melts}.
\newblock \emph{\bibinfo{journal}{Geophys. Res. Lett.}} \textbf{\bibinfo{volume}{31}} (\bibinfo{year}{2004}).

\bibitem{HIRSCHMANN201238}
\bibinfo{author}{Hirschmann, M.}, \bibinfo{author}{Withers, A.}, \bibinfo{author}{Ardia, P.} \& \bibinfo{author}{Foley, N.}
\newblock \bibinfo{title}{Solubility of molecular hydrogen in silicate melts and consequences for volatile evolution of terrestrial planets}.
\newblock \emph{\bibinfo{journal}{Earth Planet. Sci. Lett.}} \textbf{\bibinfo{volume}{345-348}}, \bibinfo{pages}{38--48} (\bibinfo{year}{2012}).


\bibitem{guimond2023mantle}
\bibinfo{author}{Guimond, C. M.}, \bibinfo{author}{Shorttle, O.} \& \bibinfo{author}{Rudge, J. F.}
\newblock \bibinfo{title}{Mantle mineralogy limits to rocky planet water inventories}.
\newblock \emph{\bibinfo{journal}{Mon. Not. R. Astron. Soc.}} \textbf{\bibinfo{volume}{521}} (\bibinfo{year}{2023}).

\bibitem{Katz}
\bibinfo{author}{Katz, R.~F.}, \bibinfo{author}{Spiegelman, M.} \& \bibinfo{author}{Langmuir, C.~H.}
\newblock \bibinfo{title}{A new parameterization of hydrous mantle melting}.
\newblock \emph{\bibinfo{journal}{Geochem. Geophys. Geosyst.}} \textbf{\bibinfo{volume}{4}} (\bibinfo{year}{2003}).

\bibitem{rimmer2021hydroxide}
\bibinfo{author}{Rimmer, P.~B.} \emph{et~al.}
\newblock \bibinfo{title}{Hydroxide salts in the clouds of {Venus}: {Their} effect on the sulfur cycle and cloud droplet p{H}}.
\newblock \emph{\bibinfo{journal}{Planet. Sci. J.}} \textbf{\bibinfo{volume}{2}}, \bibinfo{pages}{133} (\bibinfo{year}{2021}).


\bibitem{2023NatCoHelbert}
\bibinfo{author}{{Helbert}, J.} \emph{et~al.}
\newblock \bibinfo{title}{{The second Venus flyby of BepiColombo mission reveals stable atmosphere over decades}}.
\newblock \emph{\bibinfo{journal}{Nat. Commun.}} \textbf{\bibinfo{volume}{14}}, \bibinfo{pages}{8225} (\bibinfo{year}{2023}).

\bibitem{Filiberto}
\bibinfo{author}{Filiberto, J.}, \bibinfo{author}{Trang, D.}, \bibinfo{author}{Treiman, A.~H.} \& \bibinfo{author}{Gilmore, M.~S.}
\newblock \bibinfo{title}{Present-day volcanism on {V}enus as evidenced from weathering rates of olivine}.
\newblock \emph{\bibinfo{journal}{Sci. Adv.}} \textbf{\bibinfo{volume}{6}} (\bibinfo{year}{2020}).

\bibitem{oyama1980pioneer}
\bibinfo{author}{Oyama, V.} \emph{et~al.}
\newblock \bibinfo{title}{Pioneer {V}enus gas chromatography of the lower atmosphere of {V}enus}.
\newblock \emph{\bibinfo{journal}{J. Geophys. Res.: Space Phys.}} \textbf{\bibinfo{volume}{85}}, \bibinfo{pages}{7891--7902} (\bibinfo{year}{1980}).

\bibitem{surkov1986venus}
\bibinfo{author}{{Surkov}, Y.~A.}, \bibinfo{author}{{Moskalyova}, L.~P.}, \bibinfo{author}{{Kharyukova}, V.~P.}, \bibinfo{author}{{Dudin}, A.~D.}, \bibinfo{author}{{Smirnov}, G.~G.} \& \bibinfo{author}{{Zaitseva}, S.~Y.}
\newblock \bibinfo{title}{{Venus rock composition at the {Vega} 2 landing site}}.
\newblock \emph{\bibinfo{journal}{J. Geophys. Res.: Solid Earth}} \bibinfo{pages}{E215--E218} (\bibinfo{year}{1986}).

\bibitem{2008Marcq}
\bibinfo{author}{{Marcq}, E.} \emph{et~al.}
\newblock \bibinfo{title}{{A latitudinal survey of {CO}, {OCS}, \ce{H$_2$O}, and {SO$_2$} in the lower atmosphere of {Venus}: Spectroscopic studies using {VIRTIS-H}}}.
\newblock \emph{\bibinfo{journal}{J. Geophys. Res.: Planets}} \textbf{\bibinfo{volume}{113}}, \bibinfo{pages}{E00B07} (\bibinfo{year}{2008}).

\bibitem{2017Radoman}
\bibinfo{author}{{Radoman-Shaw}, B.~G.} \emph{et~al.}
\newblock \bibinfo{title}{The stability of calcium silicates and calcium carbonate on the surface of {V}enus} 
\newblock \emph{\bibinfo{journal}{Lunar Planet. Sci. Conf. Abstr.}} 
(\bibinfo{year}{2017}).

\bibitem{berger2019experimental}
\bibinfo{author}{Berger, G.} \emph{et~al.}
\newblock \bibinfo{title}{Experimental exploration of volcanic rocks-atmosphere interaction under {V}enus surface conditions}.
\newblock \emph{\bibinfo{journal}{Icarus}} \textbf{\bibinfo{volume}{329}}, \bibinfo{pages}{8--23} (\bibinfo{year}{2019}).

\bibitem{teffeteller2022experimental}
\bibinfo{author}{Teffeteller, H.} \emph{et~al.}
\newblock \bibinfo{title}{An experimental study of the alteration of basalt on the surface of {V}enus}.
\newblock \emph{\bibinfo{journal}{Icarus}} \textbf{\bibinfo{volume}{384}}, \bibinfo{pages}{115085} (\bibinfo{year}{2022}).

\bibitem{Zolotov2019}
\bibinfo{author}{{Zolotov}, M.} \emph{\bibinfo{booktitle}{Oxford research encyclopedia of planetary science}}
\newblock \bibinfo{title}{{Chemical Weathering on {V}enus}} ((\bibinfo{publisher}{Oxford University Press}, \bibinfo{year}{2018}).

\bibitem{FEGLEY1997416}
\bibinfo{author}{Fegley, B.}, \bibinfo{author}{Zolotov, M.~Y.} \& \bibinfo{author}{Lodders, K.}
\newblock \bibinfo{title}{The oxidation state of the lower atmosphere and surface of {V}enus}.
\newblock \emph{\bibinfo{journal}{Icarus}} \textbf{\bibinfo{volume}{125}}, \bibinfo{pages}{416--439} (\bibinfo{year}{1997}).

\bibitem{1977Ohmoto}
\bibinfo{author}{{Ohmoto}, H.} \& \bibinfo{author}{{Kerrick}, D.~M.}
\newblock \bibinfo{title}{{Devolatilization equilibria in graphitic systems}}.
\newblock \emph{\bibinfo{journal}{Am. J. Science}} \textbf{\bibinfo{volume}{277}}, \bibinfo{pages}{1013--1044} (\bibinfo{year}{1977}).

\bibitem{pollack1993near}
\bibinfo{author}{Pollack, J.~B.} \emph{et~al.}
\newblock \bibinfo{title}{Near-infrared light from {Venus}' nightside: {A} spectroscopic analysis}.
\newblock \emph{\bibinfo{journal}{Icarus}} \textbf{\bibinfo{volume}{103}}, \bibinfo{pages}{1--42} (\bibinfo{year}{1993}).

\bibitem{TREIMAN2012534}
\bibinfo{author}{Treiman, A.~H.} \& \bibinfo{author}{Bullock, M.~A.}
\newblock \bibinfo{title}{Mineral reaction buffering of {Venus}' atmosphere: {A} thermochemical constraint and implications for {Venus}-like planets}.
\newblock \emph{\bibinfo{journal}{Icarus}} \textbf{\bibinfo{volume}{217}}, \bibinfo{pages}{534--541} (\bibinfo{year}{2012}).
\newblock \bibinfo{note}{Advances in Venus Science}.

\bibitem{RodriguezJ}
\bibinfo{author}{{Rodriguez}, J.~M.}, \bibinfo{author}{{Prather}, M.~J.} \& \bibinfo{author}{{McElroy}, M.~B.}
\newblock \bibinfo{title}{{Hydrogen on {Venus}: Exospheric distribution and escape}}.
\newblock \emph{\bibinfo{journal}{Planet. Space Sci.}} \textbf{\bibinfo{volume}{32}}, \bibinfo{pages}{1235--1255} (\bibinfo{year}{1984}).

\bibitem{HodgesH}
\bibinfo{author}{{Hodges}, J., Richard~R.} \& \bibinfo{author}{{Tinsley}, B.~A.}
\newblock \bibinfo{title}{{The influence of charge exchange on the velocity distribution of hydrogen in the {Venus} exosphere}}.
\newblock \emph{\bibinfo{journal}{J. Geophys. Res.}} \textbf{\bibinfo{volume}{91}}, \bibinfo{pages}{13649--13658} (\bibinfo{year}{1986}).

\bibitem{BraceL}
\bibinfo{author}{{Brace}, L.~H.} \emph{et~al.}
\newblock \bibinfo{title}{{The ionotail of {Venus}: {Its} configuration and evidence for ion escape}}.
\newblock \emph{\bibinfo{journal}{J. Geophys. Res.}} \textbf{\bibinfo{volume}{92}}, \bibinfo{pages}{15--26} (\bibinfo{year}{1987}).

\bibitem{HuntenD}
\bibinfo{author}{{Hunten}, D.~M.}, \bibinfo{author}{{Donahue}, T.~M.}, \bibinfo{author}{{Walker}, J.~C.~G.} \& \bibinfo{author}{{Kasting}, J.~F.}
\newblock \emph{\bibinfo{booktitle}{Origin and Evolution of Planetary and Satellite Atmospheres}} \bibinfo{title}{ Chapter \textit{{Escape of atmospheres and loss of water.}}} \bibinfo{pages}{386--422} (\bibinfo{publisher}{University of Arizona Press}, \bibinfo{year}{1989}).

\bibitem{DonahueT}
\bibinfo{author}{{Donahue}, T.~M.} \& \bibinfo{author}{{Hartle}, R.~E.}
\newblock \bibinfo{title}{{Solar cycle variations in H$^{+}$ and d$^{+}$ densities in the {Venus} ionosphere: Implications for escape}}.
\newblock \emph{\bibinfo{journal}{Geophys. Res. Lett.}} \textbf{\bibinfo{volume}{19}}, \bibinfo{pages}{2449--2452} (\bibinfo{year}{1992}).

\bibitem{LAMMER20061445}
\bibinfo{author}{Lammer, H.} \emph{et~al.}
\newblock \bibinfo{title}{Loss of hydrogen and oxygen from the upper atmosphere of {Venus}}.
\newblock \emph{\bibinfo{journal}{Planet. Space Sci.}} \textbf{\bibinfo{volume}{54}}, \bibinfo{pages}{1445--1456} (\bibinfo{year}{2006}).
\newblock \bibinfo{note}{The Planet {V}enus and the {V}enus {E}xpress {M}ission}.

\bibitem{peucker1996accretion}
\bibinfo{author}{Peucker-Ehrenbrink, B.}
\newblock \bibinfo{title}{Accretion of extraterrestrial matter during the last 80 million years and its effect on the marine osmium isotope record}.
\newblock \emph{\bibinfo{journal}{Geochim. Cosmochim. Acta}} \textbf{\bibinfo{volume}{60}}, \bibinfo{pages}{3187--3196} (\bibinfo{year}{1996}).

\bibitem{EBEL2000339}
\bibinfo{author}{Ebel, D.~S.} \& \bibinfo{author}{Grossman, L.}
\newblock \bibinfo{title}{Condensation in dust-enriched systems}.
\newblock \emph{\bibinfo{journal}{Geochim. Cosmochim. Acta}} \textbf{\bibinfo{volume}{64}}, \bibinfo{pages}{339--366} (\bibinfo{year}{2000}).

\bibitem{kerrick2001metamorphic}
\bibinfo{author}{Kerrick, D.} \& \bibinfo{author}{Connolly, J.}
\newblock \bibinfo{title}{Metamorphic devolatilization of subducted marine sediments and the transport of volatiles into the {E}arth's mantle}.
\newblock \emph{\bibinfo{journal}{Nature}} \textbf{\bibinfo{volume}{411}}, \bibinfo{pages}{293--296} (\bibinfo{year}{2001}).

\bibitem{2022Junxing}
\bibinfo{author}{{Chen}, J.} \emph{et~al.}
\newblock \bibinfo{title}{{Venus' light slab hinders its development of planetary-scale subduction}}.
\newblock \emph{\bibinfo{journal}{Na. Comm.}} \textbf{\bibinfo{volume}{13}}, \bibinfo{pages}{7647} (\bibinfo{year}{2022}).

\bibitem{2020Lour}
\bibinfo{author}{{Louren{\c{c}}o}, D.~L.}, \bibinfo{author}{{Rozel}, A.~B.}, \bibinfo{author}{{Ballmer}, M.~D.} \& \bibinfo{author}{{Tackley}, P.~J.}
\newblock \bibinfo{title}{{Plutonic-Squishy Lid: A New Global Tectonic Regime Generated by Intrusive Magmatism on Earth-Like Planets}}.
\newblock \emph{\bibinfo{journal}{Geochem. Geophys. Geosyst.}} \textbf{\bibinfo{volume}{21}}, \bibinfo{pages}{e08756} (\bibinfo{year}{2020}).

\bibitem{stofan1992}
\bibinfo{author}{Stofan, E.~R.} \emph{et~al.}
\newblock \bibinfo{title}{Global distribution and characteristics of coronae and related features on {V}enus: Implications for origin and relation to mantle processes}.
\newblock \emph{\bibinfo{journal}{J. Geophys. Res.: Planets}} \textbf{\bibinfo{volume}{97}}, \bibinfo{pages}{13347--13378} (\bibinfo{year}{1992}).

\bibitem{Barsukov1992}
\bibinfo{author}{Barsukov, V.~L.}, \bibinfo{author}{Basilevsky, A.~T.}, \bibinfo{author}{Volkov, V.~P.} \& \bibinfo{author}{Zharkov, V.~N.}
\newblock \emph{\bibinfo{booktitle}{Venus geology, geochemistry and geophysics: research results from the {USSR}}} \bibinfo{pages}{386--422} (\bibinfo{publisher}{University of Arizona Press}, \bibinfo{year}{1992}).

\bibitem{Surkovvenera}
\bibinfo{author}{Surkov, Y.~A.}, \bibinfo{author}{Barsukov, V.~L.}, \bibinfo{author}{Moskalyeva, L.~P.}, \bibinfo{author}{Kharyukova, V.~P.} \& \bibinfo{author}{Kemurdzhian, A.~L.}
\newblock \bibinfo{title}{New data on the composition, structure, and properties of {V}enus rock obtained by {V}enera 13 and {V}enera 14}.
\newblock \emph{\bibinfo{journal}{J. Geophys. Res.: Solid Earth}} \textbf{\bibinfo{volume}{89}}, \bibinfo{pages}{B393--B402} (\bibinfo{year}{1984}).

\bibitem{Herrick2023}
\bibinfo{author}{Herrick, R.~R.} \& \bibinfo{author}{Hensley, S.}
\newblock \bibinfo{title}{Surface changes observed on a {Venusian} volcano during the magellan mission}.
\newblock \emph{\bibinfo{journal}{Science}}  (\bibinfo{year}{2023}).

\bibitem{fegley1989estimation}
\bibinfo{author}{Fegley, B.~J.} \& \bibinfo{author}{Prinn, R.~G.}
\newblock \bibinfo{title}{Estimation of the rate of volcanism on {Venus} from reaction rate measurements}.
\newblock \emph{\bibinfo{journal}{Nature}}  (\bibinfo{year}{1989}).

\bibitem{GrinspoonDH}
\bibinfo{author}{{Grinspoon}, D.~H.}
\newblock \bibinfo{title}{{Implications of the high {D/H} ratio for the sources of water in {Venus}' atmosphere}}.
\newblock \emph{\bibinfo{journal}{Nature}} \textbf{\bibinfo{volume}{363}}, \bibinfo{pages}{428--431} (\bibinfo{year}{1993}).

\bibitem{turbet2023water}
\bibinfo{author}{Turbet, M.} \emph{et~al.}
\newblock \bibinfo{title}{Water condensation zones around main sequence stars}.
\newblock \emph{\bibinfo{journal}{Astron. Astrophys.}}  (\bibinfo{year}{2023}).

\bibitem{Kane2019}
\bibinfo{author}{Kane, S.} \emph{et~al.}
\newblock \bibinfo{title}{Venus as a laboratory for exoplanetary science}.
\newblock \emph{\bibinfo{journal}{J.of Geophys. Res.: Planets}} \textbf{\bibinfo{volume}{124}} (\bibinfo{year}{2019}).









\bibitem{emcee}
\bibinfo{author}{{Foreman-Mackey}, D.} \emph{et~al.}
\newblock \bibinfo{title}{{emcee: The {MCMC} Hammer}}.
\newblock \bibinfo{howpublished}{Astrophysics Source Code Library, record ascl:1303.002} (\bibinfo{year}{2013}).

\bibitem{liggins2020can}
\bibinfo{author}{Liggins, P.,}, \bibinfo{author}{Shorttle, O.,}, \& \bibinfo{author}{Rimmer, P. B.} (\bibinfo{year}{2020}).
\newblock \bibinfo{title}{Can volcanism build hydrogen-rich early atmospheres?}.
\newblock \emph{\bibinfo{journal}{Earth Planet. Sci. Lett.}}, \textbf{\bibinfo{volume}{550}}, \bibinfo{pages}{116546}.

\bibitem{Liggins2022}
\bibinfo{author}{Liggins, P.}, \bibinfo{author}{Jordan, S.}, \bibinfo{author}{Rimmer, P.~B.} \& \bibinfo{author}{Shorttle, O.}
\newblock \bibinfo{title}{Growth and evolution of secondary volcanic atmospheres: {I}. {Identifying} the geological character of hot rocky planets}.
\newblock \emph{\bibinfo{journal}{J. of Geophys. Res.: Planets}} \textbf{\bibinfo{volume}{127}} (\bibinfo{year}{2022}).

\bibitem{Rimmer2016}
\bibinfo{author}{Rimmer, P.~B.} \& \bibinfo{author}{Helling, C.}
\newblock \bibinfo{title}{A chemical kinetics network for lightning and life in planetary atmospheres}.
\newblock \emph{\bibinfo{journal}{Astrophys. J. Suppl. Ser.}} \textbf{\bibinfo{volume}{224}} (\bibinfo{year}{2016}).

\bibitem{hobbs2021sulfur}
\bibinfo{author}{Hobbs, R.}, \bibinfo{author}{Rimmer, P.B.}, \bibinfo{author}{Rimmer, P. B.},\bibinfo{author}{Shorttle, O.} \& \bibinfo{author}{Madhusudhan, N.}
\newblock \bibinfo{title}{Sulfur chemistry in the atmospheres of warm and hot {Jupiters}}.
\newblock \emph{\bibinfo{journal}{Mon. Not. R. Astron. Soc.}} \textbf{\bibinfo{volume}{506}} (\bibinfo{year}{2021}).


\bibitem{fastchem}
\bibinfo{author}{Stock, Joachim W.}, \bibinfo{author}{Kitzmann, Daniel} \& \bibinfo{author}{Patzer, A. Beate C.}
\newblock {\bibinfo{title}{FastChem 2: An improved computer program to determine the gas-phase chemical equilibrium composition for arbitrary element distributions}}.
\newblock {\em \bibinfo{journal}{Mon. Not. R. Astron. Soc.}}
\newblock {\textbf{\bibinfo{volume}{517}}},
\newblock \bibinfo{pages}{4070--4080} (\bibinfo{year}{2022}).

\bibitem{Liggins2022b}
\bibinfo{author}{Liggins, P.}, \bibinfo{author}{Jordan, S.}, \bibinfo{author}{Rimmer, P.~B.} \& \bibinfo{author}{Shorttle, O.}
\newblock \bibinfo{title}{Growth and evolution of secondary volcanic atmospheres: 2. {The} importance of kinetics}.
\newblock \emph{\bibinfo{journal}{J. Geophys. Res.: Planets}} \textbf{\bibinfo{volume}{128}} (\bibinfo{year}{2023}).

\bibitem{woitke2018equilibrium}
\bibinfo{author}{Woitke, P.} \emph{et~al.}
\newblock \bibinfo{title}{Equilibrium chemistry down to 100 K: Impact of silicates and phyllosilicates on the carbon to oxygen ratio}.
\newblock \emph{\bibinfo{journal}{Astron. Astrophys.}}, \textbf{\bibinfo{volume}{614}}, \bibinfo{pages}{A1}  (\bibinfo{year}{2018}).

\bibitem{rimmer2019hydrogen}
\bibinfo{author}{Rimmer, P. B.}  \& \bibinfo{author}{Rugheimer, S.}
\newblock \bibinfo{title}{Hydrogen cyanide in nitrogen-rich atmospheres of rocky exoplanets}.
\newblock \emph{\bibinfo{journal}{Icarus}}, \textbf{\bibinfo{volume}{329}}, \bibinfo{pages}{124--131}  (\bibinfo{year}{2019}).

\bibitem{bezos2005fe3}
\bibinfo{author}{B{\'e}zos, A.} \& \bibinfo{author}{Humler, E.}
\newblock \bibinfo{title}{The Fe3+/$\Sigma$Fe ratios of MORB glasses and their implications for mantle melting}.
\newblock \emph{\bibinfo{journal}{Geochim. Cosmochim. Acta}}, \textbf{\bibinfo{volume}{69}}, \bibinfo{pages}{711--725}  (\bibinfo{year}{2005}).



\bibitem{ANDERSON1968}
\bibinfo{author}{Anderson, A.~D.}
\newblock \bibinfo{title}{{Superadiabatic Atmospheric Layer on {Venus}, as inferred from the {Venera}-4 Probe Measurements}}.
\newblock \emph{\bibinfo{journal}{Nature}} \textbf{\bibinfo{volume}{217}}, \bibinfo{pages}{627--628} (\bibinfo{year}{1968}).

\bibitem{OyamaPIONEER}
\bibinfo{author}{Oyama, V.~I.} \emph{et~al.}
\newblock \bibinfo{title}{Pioneer {Venus} gas chromatography of the lower atmosphere of {Venus}}.
\newblock \emph{\bibinfo{journal}{J. Geophys. Res.: Space Phys.}} \textbf{\bibinfo{volume}{85}}, \bibinfo{pages}{7891--7902} (\bibinfo{year}{1980}).

\bibitem{KRASNOPOLSKY2008377}
\bibinfo{author}{Krasnopolsky, V.~A.}
\newblock \bibinfo{title}{High-resolution spectroscopy of {Venus}: {Detection} of {OCS}, upper limit to {H$_2$S}, and latitudinal variations of {CO and HF }in the upper cloud layer}.
\newblock \emph{\bibinfo{journal}{Icarus}} \textbf{\bibinfo{volume}{197}}, \bibinfo{pages}{377--385} (\bibinfo{year}{2008}).


\bibitem{Greaves2021}
\bibinfo{author}{Greaves, J.~S.} \emph{et~al.}
\newblock \bibinfo{title}{{Phosphine gas in the cloud decks of {Venus}}}.
\newblock \emph{\bibinfo{journal}{Nat. Astron.}} \textbf{\bibinfo{volume}{5}}, \bibinfo{pages}{655--664} (\bibinfo{year}{2021}).

\end{thebibliography}
\end{document}